\documentclass{eas}
\usepackage{natbib}
\usepackage{graphicx}
\usepackage{journals}
\PassOptionsToPackage{linktocpage}{hyperref}
\newcommand{\K}{{\rm K}}

\def\gsim{\;\rlap{\lower 2.5pt
 \hbox{$\sim$}}\raise 1.5pt\hbox{$>$}\;}
\def\lsim{\;\rlap{\lower 2.5pt
   \hbox{$\sim$}}\raise 1.5pt\hbox{$<$}\;}

\newcommand{\HI}{H$\,$\textsc{i}}

\newcommand{\CIII}{C$\,$\textsc{iii}}
\newcommand{\CIV}{C$\,$\textsc{iv}}
\newcommand{\NV}{N$\,$\textsc{v}}
\newcommand{\OV}{O$\,$\textsc{v}}
\newcommand{\OVI}{O$\,$\textsc{vi}}
\newcommand{\SiIII}{Si$\,$\textsc{iii}}
\newcommand{\SiIV}{Si$\,$\textsc{iv}}

\newcommand{\lya}{Ly$\alpha$}

\begin{document}

\title{How did the IGM become enriched?}
\author{Anthony Aguirre}\address{Department of Physics, UC Santa Cruz,
1156 High St., Santa Cruz CA, 95064, USA}
\author{Joop Schaye}\address{Leiden Observatory, Leiden University,
  P.O. Box 9513, 2300 RA Leiden, The Netherlands}
%
%
\begin{abstract}

The enrichment of the intergalactic medium with heavy elements is a
process that lies at the nexus of poorly-understood aspects of
physical cosmology.  We review current understanding of the processes
that may remove metals from galaxies, the basic predictions of these
models, the key observational constraints on enrichment, and how
intergalactic enrichment may be used to test cosmological simulations.

\end{abstract}
\maketitle
\section{Introduction}

The widespread existence of metals outside of galaxies has been known
for a decade (e.g., Cowie et al.\ 1995) by their absorption lines in
high-$z$ QSO absorption spectra, and was quite surprising when
discovered.  It has since become clear that this enrichment is related
to many other aspects of physical cosmology, and that the question of
``how did the intergalactic medium become enriched?" has a number of
components that we would like to understand.  Among them:

\begin{itemize}
\item{The low-density ($\delta \equiv \rho/\left <\rho\right > - 1 \la
  10$) intergalactic medium (IGM), as 
probed by the \lya\ forest and through \CIII, \CIV, \SiIII, \SiIV,
\OV, \OVI, and other transitions, is at least partly enriched at all
  redshifts and densities
probed~\citep[e.g.,][]{1996AJ....112..335S,1998Natur.394...44C,2000AJ....120.1175E,2000ApJ...541L...1S,2001ApJ...561L.153S,2002ApJ...578...43C,2002ApJ...579..500T,2002A&A...396L..11B,2003ApJ...596..768S,2003astro.ph..7557B,2003ApJ...594..695P,2004ApJ...602...38A,2004A&A...419..811A,2004ApJ...606...92S,2004ApJ...617..718P,2006MNRAS.371L..78R}. How
did these metals get where they are?}

\item{The intracluster medium is highy enriched to $Z\sim
0.2-0.5\,Z_\odot$, with most metals produced in clusters residing in
the intracluster medium rather than in the galaxies
themselves~\citep[e.g.,][]{1997ApJ...488...35R}.  How did such
efficient metal ejection occur?}
 
\item{Semi-analytic and numerical models of galaxy formation
overpredict the stellar masses of (in particular small) galaxies
unless they incorporate significant feedback such as would drive
material from
galaxies~\citep[e.g.,][]{1993MNRAS.264..201K,1999MNRAS.310.1087S}. What
is the relation between enrichment and galaxy formation?}

\item{The $z\sim 0$ (and $z\sim 2$) mass-metallicity relation is
indicative of loss of metals by
galaxies~\cite[e.g.,][]{2004ApJ...613..898T,2006ApJ...644..813E}. How
much baryonic mass and metal mass is ejected by galaxies as they
form?}
 
\item{Some galaxies at $z \sim ~ 0$~\citep{2001ASPC..240..345H}, and
nearly all galaxies at $z > 2$~\cite[e.g.,][]{2003ApJ...584...45A},
are observed to drive winds that might escape.  Do they escape?  What
is their impact?  }

\end{itemize}

In connection with these facts, a general picture has emerged that
galactic winds -- driven largely from young and/or starburst galaxies
-- have enriched the IGM both within clusters and the general field.
The same feedback may account for the dearth of low-luminosity
galaxies (relative to the halo mass function), and also the
mass-metallicity relationship of galaxies.  However, a detailed
understanding of the various feedback processes is lacking, and there
are still open questions, and controversies, concerning the time
at which various components of enrichment occurred, exactly how, and
what this tells us about galaxy formation.

The rest of this article will review recent progress in piecing
together this picture.  First we schematically review the physics
behind outflows, the basic ways of assessing the predictions of these
for the observed intergalactic (IG) enrichment, and what general
predictions seem to be agreed-upon.  We then discuss how the
metallicity of the low-density IGM is measured, and the basic results
obtained using these techniques.  Finally, we discuss what is probably
the best way to get a true handle on the enrichment mechanism: direct
comparison between detailed feedback simulations and the large data
sets now becoming available.

\section{Basic physics of galactic winds}
\label{sec-models}
While a
single supernova explosion cannot blow material out of any but the
smallest galaxy, a large set of contemporaneous and co-spatial
explosions might.  In this case, the overlap of super-heated bubbles
can overlap and merge into a ``superbubble" that blows out of the
galaxy and forms an outflowing large-scale
wind~\cite[e.g.,][]{1990ApJS...74..833H}.  A 
common way to model this has been to assume that the superbubble will
form a ``supershell" for which equations of motion can be evolved that
include the driving pressure, gravity, the ram pressure of the medium
encountered,
etc.~\cite[e.g.,][]{2001ApJ...561..521A,2001MNRAS.321..450T,2005MNRAS.359.1201B}. Even
if the shell fragments (as is expected), another shell might form, and
the evolution can be followed by encompassing repeated
shell-fragmentation in the form of an ``entrainment fraction"
specifying how much of the ambient material must be carried by
the outflow along with the superheated gas.

A second basic mechanism has more recently been proposed
by~\cite{2005ApJ...618..569M}, in which it is momentum deposition by
radiation pressure (coupled via dust to the gas\footnote{If the dust
is somehow {\em not} coupled to the gas, then an outflow of dust could
ensue and contribute to IG enrichment;
see~\citet{2001ApJ...556L..11A,2005MNRAS.358..379B}.}), or cosmic ray
pressure~\citep{2006astro.ph..9796S}, that drives the outflow.

These two possible mechanisms may coexist and would both result in a
rather complicated outflow structure.  A first-order model of the
effect of these outflows on the host galaxy and the nearby IGM
can, however, be generated by assuming that whatever the outflow does
at very small radii, it eventually organizes into a wind with some mass
outflow rate $\dot m_{\rm sfr}$, velocity $V_{\rm wind}$, and
entrainment fraction, and does so in such a way as to conserve some
quantity (energy or momentum).  The
two mechanisms differ in the suggested relation between, in
particular, $\dot m_{\rm out}$ and $V_{\rm wind}$.  For energy driven
winds, it has been argued that $V_{\rm wind}$ should be roughly
independent of the progenitor galaxy
mass~\cite[e.g.,][]{1990ApJS...74..833H}.  Then if the feedback
efficiency is also fixed, equating the energy generation to the energy
carried by the wind indicates that: $$\dot m_{\rm out} \propto \dot
m_{\rm sfr}, \ \ V_{\rm wind}\simeq const.$$

For winds driven by radiation pressure, it is more natural for both
the wind speed and outflow rate to depend on the progenitor galaxy
mass (or velocity dispersion $\sigma$), giving a wind possible wind
prescription at large radii
of~\citep{2005ApJ...618..569M,2006astro.ph..5651O}:
$$ \dot m_{\rm out} \propto \dot m_{\rm sfr}/\sigma, \ \ V_{\rm
wind}\simeq 3\sigma,
$$ with the wind momentum outflow $\dot m_{\rm out}V_{\rm wind}$
equated to a fixed fraction of the momentum input (i.e. luminosity) of
the source.

While these are two interesting perspectives to take, a very great
deal happens between an erg of energy (or g cm/s of momentum) being
deposited, and the wind several kpc away.  In particular, while energy
and momentum are certainly conserved in these systems, winds may
radiate much of their energy and much of the momentum may cancel
out if, as in starbursts, it is generated by a distribution of
sources.  

That being said, suppose we have some prescription for the wind
parameters as a function of galaxy properties.  How might we make
predictions for intergalactic enrichment?

\section{Modeling galactic winds}

Two basic approaches are used in making predictions for IG enrichment
from a given wind model: the semi-analytic, and the numerical.

Semi-analytic
approaches~\citep[e.g.,][]{2001ApJ...561..521A,2001ApJ...555...92M,2002ApJ...574..590S,2002ApJ...581..836T,2003ApJ...588...18F,2005MNRAS.358..379B,2005MNRAS.359.1201B,2006astro.ph..6423P}
generally proceed as follows:

\begin{enumerate}
\item A cosmological simulation is used to determine galaxy locations,
halo masses, and gravitational potentials.

\item The simulations outputs themselves (for hydrodynamic
simulations) or semi-analytic prescriptions overlaid on the halos (for
dark matter simulations) are used to determine galaxy star-formation
rates, metallicities, gas properties, inflow rates, etc.

\item A prescription based on a simplified wind model is applied to
the galaxy properties to determine how far a wind would propagate
during a ``time step" and how the expelled metals would be
distributed.  These metals are then ``painted on" to the gas or dark
particles.

\item The metallicity of particles is then tracked, to yield the
metallicity distribution at later cosmological epochs.

\end{enumerate}

In the numerical
approach~\citep[e.g.,][]{1999ApJ...519L.109C,2002ApJ...578L...5T,2002ApJ...581..836T,2005ApJ...620L..13A,2005ApJ...635...86C,2006astro.ph..5651O},
one instead:

\begin{enumerate}

\item Directly simulates gas physics anew for each trial.

\item Uses prescriptions for star formation, feedback of both energy
and metals into the gas surrounding star-forming regions, and possibly
for the launching of winds.\footnote{Of course, in principle, the wind
-- if driven by energy injection -- should launch itself.  That this
generally does not occur without some sort of assistance is generally
attributed to insufficient resolution.}

\item Tracks metals accumulating in the gas particles, and potentially
uses these metallicities in the cooling rates of the gas.

\end{enumerate}

Relative to the numerical method, the primary advantage of the
semi-analytic method is speed: many realizations can be computed
on a desktop computer in the time that one simulation can be run on a
supercomputer. Another advantage is that semi-analytic methods allow
a clearer connection to be made between the physical ingredients of
the model and the results, possibly leading to a better understanding of
the actual physics that is responsible for particular effects.

Numerical methods, in contrast, treat the gas physics {\em much} more
accurately and can, for example, assess the impact of feedback on the
thermal state of the gas.  Depending on the resolving power of the
simulations, fewer prescriptions for physics need to be used than in the
semi-analytic method, since the physics can be directly simulated.
Simulations naturally allow a self-consistent treatment of the effect
of feedback on star formation.  Finally, they easily allow for the
generation of mock observations.

There is a fairly large literature on IG metals using both approaches,
so we here focus on what appear to be general and (fairly) well-agreed
upon findings:

\begin{itemize}

\item Galaxies at $z > 2$ essentially all drive winds: Whether winds
are launched ``automatically" or via a prescription, if the mechanism
is consistent with winds from low-$z$ starbust galaxies, then it will also
predict that most high-$z$ galaxies drive them, since the specific SFRs are
so much higher then.  

\item Wind propagation and escape is quite sensitive to the
entrainment fraction and to $v_{\rm wind}$.  This occurs because the
two primary forces limiting wind propagation are the galaxy's
potential well (if the wind speed is not much higher than the escape
velocity) and the ram pressure of the (potentially infalling) gas that
must be swept up even if the wind is fast. Moreover, if entrainment is
significant, then the mass over which the wind energy and momentum
must be shared may be much greater.

\item The metallicity of the IGM resulting from enrichment by galactic
winds is highly inhomogeneous and probably has a filling factor $\la
10\%$.

\item Higher-redshift enrichment tends to lead to more
homogeneous enrichment at lower-redshift.  This is somewhat
contentious and depends in detail on the assumptions made, but appears
to be borne out by numerical calculations.  It occurs because high-$z$
galaxies are relatively smaller and more numerous, and because the winds
have more time to propagate.  It appears to occur in spite of the fact
that the earliest galaxies of any given mass tend to form in highly
biased and highly-clustered regions.

\item There is a strong correlation between the gas metallicity and
the gas density, which follows simply from the fact that galaxies form
preferentially in high-density regions.

\end{itemize}

The numerical treatments of IG enrichment reinforce many of these
conclusions.  In addition they seem to generically suggest that:

\begin{itemize}

\item A significant fraction of the metals ejected end up in hot ($\ga
10^5\,$K) gas, especially when metal-line cooling is not included, but
also even when it is.

\item Except at the highest redshifts and with the smallest sources,
winds tend to preferentially propagate outside and in-between
filaments, and also have a low filling factor, so that there is little
overall disruption of the structure or statistical description of the
Ly$\alpha$ forest itself.

\end{itemize}

\section{Assessing metals spectroscopically}

Nearly everything we know about the enrichment of the IGM at $z \ga 2$
is via optical \CIV, \CIII, \SiIV, \SiIII, \OV, and \OVI\
absorption-line spectroscopy of distant quasars.  Below we briefly
review how these assessments are performed, and the basic results
obtained to date.

\subsection{Techniques}

The standard technique is to fit line profiles to a
set of absorption features to derive (where the line is not too
saturated) column densities and (where the line is resolved)
line-widths.  This technique works well in the IGM for
relatively high-column lines where there is relatively little
contamination.  Given extremely good data, this can probe gas
densities at only a few times the cosmic
mean~\citep{2004ApJ...606...92S}.  Once determined, the ionic column
densities can might be converted into metallicities as described
below.  The prime advantages of line-fitting over the pixel-based
techniques described next are that it (a) is familiar and
straightforwardly interpreted, (b) treats high column-density lines
well, and (c) gives information on line widths.  Its main disadvantages are
that it (a) is unsuitable for statistical searches for absorption that
is weak compared with the noise or contamination; (b) is subjective in
the sense that decompositions in Voigt profiles are non-unique; (c) is
slow and dependent on prior identification of lines, making it unsuitable for
application to large datasets and simulations. 

The pixel optical depth (POD)
method~\citep{1998Natur.394...44C,1998AJ....115.2184S,1999ApJ...520..456E,2000AJ....120.1175E,2000ApJ...541L...1S,2002ApJ...576....1A,2002ApJ...579..500T,2003ApJ...596..768S,2004A&A...419..811A,2004ApJ...602...38A,2004MNRAS.347..985P,2005AJ....130.1996S,2006ApJ...638...45P}
is a complementary approach that has certain advantages, particularly
when applied to weak, widespread absorption that may be easily
confused with contaminating lines or when applied to large date sets
or simulations.

The key steps in the POD search are:
\begin{enumerate} 
\item Compute arrays of apparent pixel optical depths, $\tau_{\rm
app}(z) = -\ln{F(z) / F_{\rm cont}(z)}$, where $F$ and $F_{\rm cont}$
are the transmitted flux in the pixel and the local continuum,
respectively. Sophisticated methods have been
developed~\citep{2002ApJ...576....1A,2003ApJ...596..768S} to correct
for contamination, noise, continuum fitting errors, etc. in these
``apparent" optical depths, taking advantage of the fact that many
ions generate multiplets (e.g., \HI) or doublets (e.g., \CIV, \NV,
\OVI, and \SiIV), with known optical depth ratios.

\item Choose a ``base'' and a ``target'' transition and bin the pixel
  pairs according to the recovered POD of the former (e.g., \CIV\ as a
  function of \HI).  The combinations that have so far proved to be
  most useful are \CIV(\HI)~\citep{1998Natur.394...44C},
  \OVI(\HI)~\citep{2000ApJ...541L...1S},
  \SiIV(\CIV)~\citep{2004ApJ...602...38A},
  \CIII(\CIV)~\citep{2003ApJ...596..768S}, and
  \SiIII(\SiIV)~\citep{2004ApJ...602...38A}.
  
\item Compute the median (or any other percentile) of the target
  transition POD as a function of the binned base transition POD. A
  correlation reflects a detection.  The method can be generalized to
  measure the full distribution of target optical depths by
  simultaneously measuring multiple
  percentiles~\citep{2003ApJ...596..768S}.

\end{enumerate}

While elemental abundances are desired, we can only measure ion
abundances, whether from optical depth ratios or column density
ratios.  This is unfortunate in that uncertainties in the ionization
balance are currently the limiting factor in the study of
intergalactic abundances. But it is also good news, because it means
that we can constrain the physical conditions in the gas, since the
ionization balance depends in general on the radiation field, the gas
density, and the gas temperature.

The radiation field (if uniform), can be described in terms of a
normalization (in the form of the \HI\ ionization rate) and a spectral
shape (determined by the type of sources, and the transfer of
radiation through 
the IGM).  The UV background (UVB) models
of~\cite{2001cghr.confE..64H} have become fairly standard, (which is
not to say they may not be quite incorrect!) and provide a useful
benchmark.

In photo-ionization equilibrium, the gas density is closely related to
  the \HI\ optical depth: $n_{\rm 
  HI} \propto \rho^2 \Gamma_{\rm HI}^{-1}$, and the \HI\ ionization
  rate $\Gamma_{\rm HI}$ can be measured. A similar (but more
  detailed) relation can be obtained using mock spectra drawn from
  hydrodynamics simulations.  This relation can be used both to
  perform ionization corrections, and to indicate the density of gas
  being probed.

If the gas is very hot and/or dense, then collisional ionization may
  dominate over photo-ionization, in which case the ionization balance
  will depend only on the temperature. If photo-ionization dominates,
  then the ion fractions are still weakly dependent on the temperature
  because the recombination rates are. However, as long as $T\sim
  10^4~\K$, as is usually assumed, the temperature is not the main
  source of uncertainty in the analysis.

\subsection{Observations of IG metals at $z > 2$}

Using the spectral data and ionization-correction modeling just
discussed, in the past decade or so we have assembled a reasonably
detailed assessment of IG enrichment -- albeit with some large
outstanding questions.

Basic results thus far include:

\begin{itemize}
\item The carbon abundance (probed via \CIV) is inhomogeneous: at
fixed overdensity $\delta$ and redshift $z$, the p.d.f. for [C/H] is a
gaussian of width $\sim $0.5-1.0 dex, i.e. the 
distribution is
lognormal~\citep{2003ApJ...596..768S,2004ApJ...606...92S}. Similar
results hold for oxygen as probed via \OVI~\citep{2004ApJ...606...92S}.

\item The median carbon abundance [C/H] (and also the width of the
p.d.f.) increases with density~\citep{2003ApJ...596..768S}, and there is
(some) carbon in (some) underdense gas (corresponding to $N_{\rm HI}
\la 3\times 10^{13}$)~\citep{2003ApJ...596..768S}.

\item For a standard choice of the UV background (HM01 Q+G UVB), the
median [C/H] does not evolve from $z\sim 4$ to
$z\sim 2$~\citep{2003ApJ...596..768S}. In addition, metals exist at $z
\ga
5$~\citep{2001ApJ...561L.153S,2003ApJ...594..695P,2006MNRAS.371L..78R,2006astro.ph..5710S}.

\item The IGM is alpha-enhanced relative to the sun: [Si/C] $\sim
0.2-1.5$~\citep{2004ApJ...602...38A}, [O/C] $\ga
0.0$~\citep{2002ApJ...579..500T,2004ApJ...606...92S}. This suggests
enrichment by
predominantly Type II supernovae, perhaps even with a contribution by
hypernovae~\citep[see, e.g.,][and other contributions to this
volume]{2005ApJ...623...17Q}.
 
\item Results are UVB-dependent. A \cite{2001cghr.confE..64H} UVB
model with galaxies and quasar light included, and 10\% escape
fraction from galaxies seems plausible; a
much harder or softer does UVB does
not~\citep{2003ApJ...596..768S,2004ApJ...602...38A,2004ApJ...606...92S}
 
\item Using detected \SiIII\ and \CIII\ absorption, it can be shown
that most strong \SiIV\ and \CIV\ absorption arises in photoionized
gas~\citep{2003ApJ...596..768S,2002ApJ...576....1A}.
 
\item Metals are correlated with themselves and with galaxies
\citep{2005ApJ...629..636A,2006ApJ...638...45P,2006MNRAS.365..615S}.
 
\end{itemize}

\subsection{Basic Implications of derived abundances}

A basic implication of the above findings is that essentially all of
the results are explainable by metals ``sprinkled'' onto simulations
with no winds, at $z \ga 5$.  But does it mean that {\em all} metals
in the IGM come from such an early enrichment period?  There are some
reasons to think not.

First, a substantial fraction of all metals generated by stars at $z \ga 3$
appear to reside in the diffuse IGM:
[C/H] $\sim -2.8$, [Si/H] $\sim -2.0$, [O/H] $\sim -2.3$ (for HM01
Q+G) imply $f_{Z,IGM} ~ 10-30\%$~\citep{2006bouchebudget}.  

Second, because
their ionization fractions fall quickly with temperature above $\sim
10^5\,$K, \CIV\ and \SiIV\ cannot probe very hot gas -- metals in such
gas would simply be invisible to probes using these ions. 

In general, the {\em rough} amount of metals present in the IGM can be
reproduced by models that include galactic winds (though not with
models that do not include winds).  However, the observations
summarized above present a detailed and challenging target for the
next generation of models.  In the next section we summarize some
recent work comparing cosmological simulations -- including enrichment
-- in detail to the observed spectra.

\section{Confrontation with simulations}

While much can be inferred using observations directly, It is very
useful to compare them to synthetic spectra generated from
simulations with enrichment -- including the effect of the outflows on
the IGM.  In one
investigation of this sort,~\cite{2005ApJ...620L..13A} analyzed two
sets of high-resolution particle-pased hydrodynamic 
simulations~\citep{2003MNRAS.339..312S,2002ApJ...578L...5T} with
different feedback prescriptions that both drive strong winds at $z \gsim
2$ as per the late enrichment scenario.  Four general and qualitative
conclusions of this study were as follows.

First, the feedback simulations 
produce far too small
values of $\tau_{\rm CIV}/\tau_{\rm HI}$ and $\tau_{\rm
CIII}/\tau_{\rm CIV}$, for all UVB models employed.

Second, metal-rich gas in the simulations is too hot and too
low-density: essentially all of the intergalactic and enriched gas in
both simulations exists in low-density, high-temperature ($10^5\,K
\lsim T \lsim 10^6\,K$) bubbles.  Since both the \CIV/C ratio and the
\CIII/\CIV\ ratio fall off quickly with both increasing temperature $T
\gsim 10^5\,$K and decreasing density $\delta \lsim 10$, the gas
becomes nearly invisible in \CIV, and where \CIV\ is detected there is
virtually no accompanying \CIII.

Third, metal line cooling, which was not included, should be
important. 
If crudely modeled (by assuming that gas cools to $T\approx 10^4\,$K
if its cooling time is shorter than the Hubble time), much more \CIV\
absorption is present and the {\em median} $\tau_{\rm CIV}/\tau_{\rm
HI}$ could plausibly match the observed one.

Finally, metals nonetheless appear too low-density and too
inhomogeneous: While including metal cooling might lead to a
sufficient increase in \CIV\ absorption, it does {\em not} appear to
fix the difficulty in reproducing \CIII/\CIV.  This can be traced to
the fact that the metal-rich gas has too low density to produce the
observed ratios.  This problem may also be lessened by a proper
treatment of metal cooling, as the cooling would also affect the gas
dynamics, allowing gas to contract as it cools. But there is yet
another problem, which is that although the {\em median} $\tau_{\rm
CIV}/\tau_{\rm HI}$ can be roughly reproduced by the simulations, the
spread in $\tau_{\rm CIV}$ at a given $\tau_{\rm HI}$ cannot: the
metals in the simulations are {\em too inhomogeneous}.  This problem
seems unlikely to improve from a correct treatment of metal cooling.

In a more recent study,~\cite{2006astro.ph..5651O} have compared
simulated spectra from a large suite of feedback simulations to
observed line and pixel data. (See also contributions by Dav\'e and
Oppenheimer in the present volume for more details; here we just
mention a few points.)

The Oppenheimer \& Dav\'e simulations include metal-line cooling, and
wind prescriptions chosen to correspond to ``momentum driven" and
``energy driven" winds as discussed in Sec.~\ref{sec-models}.  In
comparison to~\cite{2005ApJ...620L..13A} there are two findings of
note. First, Dav\'e et al. find much better agreement in \CIV/\HI\ and
\CIII/\CIV\ optical depth ratios between their simulations and the data
than do~\cite{2005ApJ...620L..13A}.  This is very likely due to the
inclusion of metal line cooling.  Second, in models for which the \CIV\
optical depths are compatible with observations, the overall cosmic
carbon abundance is higher than that inferred
by~\cite{2003ApJ...596..768S} based on the same data.  This indicates
that (as argued by Dav\'e et al.) metals are being hidden in
relatively hot gas not probed by the observations.  

It would be very interesting to see how the Dav\'e et al. simulations
fare when applied to \OVI, which is much more sensitive to the
presence of hot gas, and when applied to percentiles in \CIV\ other
than the median -- which~\citet{2005ApJ...620L..13A} found difficult to
reproduce using simulations, even with cooling included.

\section{Concluding facts and opinions}

We will conclude with some facts, and some opinions.  The key facts
are:

\begin{itemize}
\item Both observations and simulations of IG enrichment are getting
drastically better, and the former are, finally, strongly constraining
the latter.

\item Comparisons between observations and simulations have to be done
  with care. 

\item Observations of IG metals strongly constrain models of feedback from
  star formation (a crucial ingredient of simulations and galaxy
  formation models). 
\end{itemize}

Some opinions:

\begin{itemize}
\item Enrichment is unlikely to be purely ``contemporary'' or
``primordial'': Enrichment almost surely occurs at $z < 4$, perhaps mostly
hidden in hot gas. However, it is unclear that $z < 6$ enrichment can
account for all measurements.

\item It seems likely that a galaxy-mass-dependent $V_{\rm wind}$ is
necessary -- but it is not clear that this requires
radiation-pressure-driven winds.

\item There is a dearth of high-quality, high-resolution simulations
of individual wind-driving galaxies in a cosmological background.
Please do some!

\end{itemize}



\end{document}